\documentclass{PoS}

\usepackage{amsmath}

\title{Towards global analysis of $b\to s\,\ell^+\ell^-$}

\ShortTitle{Towards global analysis of $b\to s\,\ell^+\ell^-$}

\author{
  \speaker{Christoph Bobeth}\\
  Institute for Advanced Study and Excellence Cluster Universe\\
  Technische Universit\"at M\"unchen\\
  D-85748 Garching, Germany \\
  E-mail: \email{christoph.bobeth@ph.tum.de}
}

\abstract{ Flavour changing neutral current decays mediated by $b\to
  s\,\ell^+\ell^-$ were under experimental investigation at $B$-factories and
  the Tevatron during the last decade and the final analysis are expected
  soon. Moreover, new data has been released this summer and more is currently
  taken at the LHC mainly by LHCb.  The theoretical methods for both inclusive
  and exclusive decays have been also refined and a global analysis of these
  decays becomes more and more feasible.  First analysis of combined $b\to
  s\,\ell^+\ell^-$ data towards a global analysis provide constraints on the
  involved short distance couplings.  Recently, specially designed observables,
  which can be determined in the angular distribution of $B\to K^*(\to K\pi)\,
  \ell^+\ell^-$, became of particular interest since they are subject of reduced
  hadronic uncertainties and currently their potential role in a global analysis
  is under investigation. 
}

\FullConference{
  The 2011 Europhysics Conference on High Energy Physics, EPS-HEP 2011,\\
  July 21-27, 2011\\
  Grenoble, Rh\^one-Alpes, France 
}

\begin{document}


Flavour physics constitutes an important part of the standard model (SM)
phenomenology.  Especially, $B$-physics at the 1st generation $B$-factories and
the Tevatron played a major role during the last decade. One of the main aims
was -- and still is -- to measure and test the picture of quark-flavour mixing
and CP violation in the SM, represented by the unitary Cabibbo-Kobayashi-Maskawa
(CKM) matrix. Great effort was put into the determination of the CKM parameters
with the help of elaborated strategies developed by the CKMfitter and UTfit
groups leading to a steadily increasing precision of the involved
parameters. Currently, this program is continued at the LHC mainly by LHCb and
in the future at the 2nd generation $B$-factories with Belle II and SuperB.

Flavour changing neutral current (FCNC) decays mediated by $b\to
s\,\ell^+\ell^-$ are not of primary importance for the determination of CKM
parameters. In the SM they are absent at tree-level, having branching fractions
of ${\cal O}(10^{-6})$ and being sensitive to contributions beyond the SM (BSM)
at the electroweak scale or even higher, depending on the BSM scenario.  They
constitute indirect probes of virtual contributions of new physics which
requires a certain degree of experimental and theoretical precision to test the
SM or to be able to claim deviations.  The experiments Belle~\cite{:2009zv}, CDF
\cite{Aaltonen:2011qs} and LHCb \cite{Blake:2011ii} started to accumulate enough
$b\to s\,\ell^+\ell^-$ events in order to begin first studies. Especially, LHCb
will provide at least a factor five more data with 2 fb$^{-1}$ until the end of
next year for exclusive final states such as $B^+ \to K^+ \,\mu^+\mu^-$ and $B^0
\to K^{*0}(\to K^+\pi^-) \,\mu^+\mu^-$.  In view of the experimental progress it
is desirable to develop strategies towards a global fit of $b\to
s\,\ell^+\ell^-$ and related FCNC decays similar to the ones existing for the
CKM matrix.

Theoretical predictions of $b\to s\, \ell^+\ell^-$ decays focus on regions of
the dilepton invariant mass~$q^2$ below and above the two narrow
$c\bar{c}$-resonances $J/\psi$ and $\psi'$, which are frequently denoted as low-
and high-$q^2$ region, respectively. In the SM, the numerically dominant
contributions in both $q^2$-regions are due to loop-induced FCNC operators,
usually denoted as ${\cal O}_{9, 10} \sim [\bar{s}\, \gamma^\mu P_L\,
b][\bar\ell\, \gamma_\mu (1, \gamma_5)\, \ell]$ for $b\to s\, \ell^+\ell^-$ and
the electric dipole operator ${\cal O}_{7} \sim m_b [\bar{s}\, \sigma^{\mu\nu}
P_R\, b] F_{\mu\nu}$ for $b \to s\gamma$. At low-$q^2$, contributions due to
$b\to s\,\bar{q}q$ ($q = u,d,s,c$) 4-quark operators are treated within QCD
factorization \cite{Beneke:2001at} based on the large recoil limit and
soft-gluon effects from $c\bar{c}$-resonances can be included following
\cite{Khodjamirian:2010vf}.  At high-$q^2$, a local expansion can be applied for
these operators due to the hard momentum $\Lambda_{\rm QCD} \ll q^2 \sim m_b^2$
of the order of the $b$-quark mass \cite{Grinstein:2004vb, Beylich:2011aq}. The
main uncertainties in predictions of exclusive decays $B\to K^*\,\gamma$, $B\to
K^{(*)} \ell^+ \ell^-$, but also others like $B_s\to \phi\,\ell^+ \ell^-$,
$\Lambda_b \to \Lambda\, \ell^+ \ell^-$, stem from form factors and lacking
sub-leading contributions in the power expansions in $\Lambda_{\rm QCD}/m_b$.

The particular kinematical limits of large and low recoil of the hadronic system
$K^{(*)}$ allow to reduce the number of form factors with the help of form
factor relations at lowest order in the above mentioned power expansions.  In
combination with the angular analysis of the 4-body final state in $B\to
K^{*}(\to K\pi)\, \ell^+ \ell^-$, which offers a large number of angular
observables $J_i(q^2)$ ($i = 1s,\, \ldots\, 9$) \cite{Kruger:1999xa}, suitable
combinations of $J_i(q^2)$ could be identified which exhibit a reduced hadronic
uncertainty and enhanced sensitivity to short-distance couplings of the SM and
BSM scenarios. At low-$q^2$ there are $A_T^{(2,3,4,5,{\rm re, im})}$
\cite{Kruger:2005ep} whereas at high-$q^2$ $H_T^{(2,3,4,5)}$
\cite{Bobeth:2010wg}. Additionally, at high-$q^2$ also combinations are known
which do not depend on the short-distance couplings \cite{Bobeth:2010wg} and
allow to probe the form factor shapes with data.  CP asymmetric combinations
with reduced hadronic uncertainties have been also found at low-$q^2$
\cite{Kruger:2005ep} and high-$q^2$ \cite{Bobeth:2011gi}.  The sensitivity to
$B_s$-mixing parameters $\phi_s$ and $\Delta \Gamma_s$ in time-integrated CP
asymmetries of $B_s \to \phi(\to K^+K^-)\,\ell^+\ell^-$ turns out to be small
\cite{Bobeth:2011gi}.  The $J_i(q^2)$ normalised to the decay rate and the
associated CP-asymmetries have been also studied model-independently and
model-dependently in great detail \cite{Bobeth:2008ij}.

\begin{figure}
\includegraphics[width=.335\textwidth]{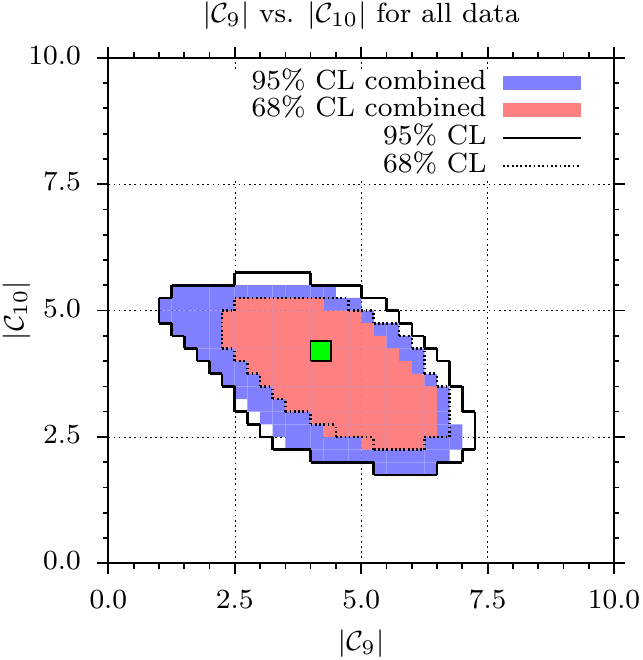}
\includegraphics[width=.325\textwidth]{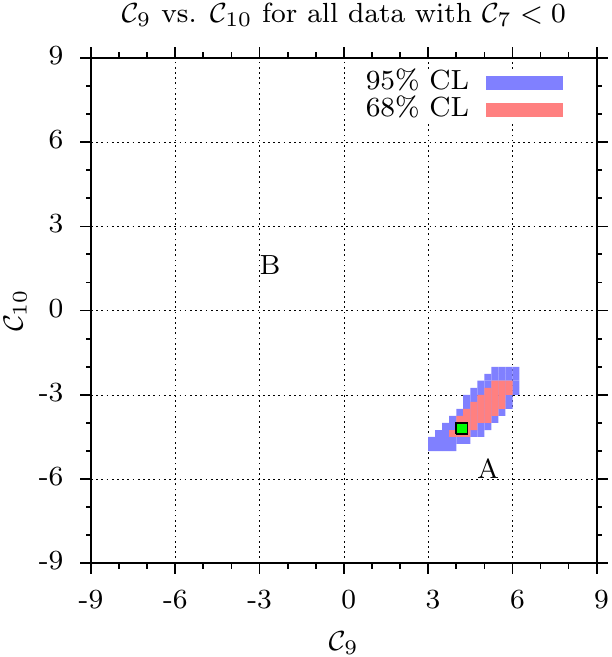}
\includegraphics[width=.325\textwidth]{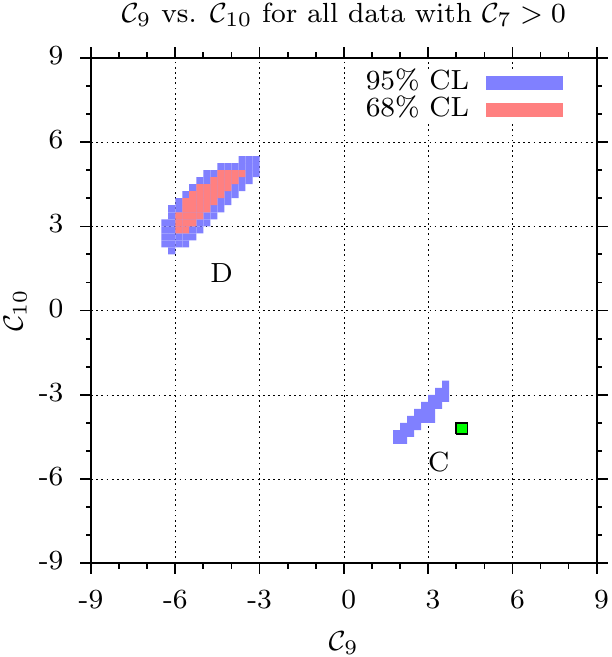}
\caption{Constraints on $|C_9|$ and $|C_{10}|$ for complex $C_{7,9,10}$ [left]
  and real $C_{9,10}$ with $C_7^{} = C_7^{\rm SM}$ [middle] as well as 
  $C_7^{} = -C_7^{\rm SM}$ [right], using low- and high-$q^2$ constraints from 
  $B\to K^{(*)}\ell^+ \ell^-$ \cite{Bobeth:2011nj}. The black contours in the
  left plot are obtained by discarding $B\to K\,\ell^+\ell^-$ data and
  the (green) square denotes the SM point.
  All observables used in \cite{Bobeth:2010wg, Bobeth:2011gi, Bobeth:2011nj} 
  are implemented in the public package \cite{EOS:website}.}
\label{fig1}
\end{figure}

The $q^2$-binning for several observables measured by Belle \cite{:2009zv}, CDF
\cite{Aaltonen:2011qs} and LHCb \cite{Blake:2011ii} falls into the low- and
high-$q^2$ regions which are accessible by theoretical methods and allows to
preform first fits of the short-distance couplings $C_{9,10}$. The constraining
potential of the combination of both $q^2$-regions has been demonstrated using
Belle and CDF $B\to K^*\ell^+\ell^-$ data from 2010 \cite{Bobeth:2010wg}.  There
the branching ratio, the lepton forward-backward asymmetry $A_{\rm FB}$ and the
longitudinal $K^*$-polarisation fraction $F_L$ in the bins $q^2 \in [1, 6]$,
$[14.18, 16.0]$ and $[> 16.0]$~GeV$^2$ has been studied for a SM operator basis
scenario with real $C_{9,10}$ and $C_7 = \pm C_7^{\rm SM}$.  In the lack of
QCD-Lattice predictions, the fits rely on extrapolations of form factors from
the low- to the high-$q^2$ region.  The according results for a scenario of
complex $C_{7,9,10}$, i.e. accounting also for CP violation beyond the SM, can
be found in \cite{Bobeth:2011gi}.  The study \cite{DescotesGenon:2011yn}
included only data from the low-$q^2$ region, but supplemented with inclusive
and exclusive $b\to s\, \gamma$ observables for the SM scenario with real
$C_{7,9,10}$ and its extension including the chirality-flipped operators ${\cal
  O}_{7', 9', 10'}$. Especially, $A_T^{(2)}$ was studied which is particularly
sensitive to $C_{7',9',10'}$, showing, that large deviations from the SM
prediction are still allowed. A very comprehensive study
\cite{Altmannshofer:2011gn} included the most recent data including also
high-$q^2$ results, except $B\to K\ell^+\ell^-$, and studies model-independent
scenarios with complex Wilson coefficients.

During this summer CDF (6.8 fb$^{-1}$) released updated results for several
exclusive $b\to s\, \ell^+\ell^-$ decays \cite{Aaltonen:2011qs} and moreover,
LHCb (309 pb$^{-1}$) presented the first results of $B\to K^*\ell^+\ell^-$
\cite{Blake:2011ii}. Figure~\ref{fig1} shows the allowed ranges of $|C_{10}|$ vs
$|C_9|$ for complex and real $C_{7,9,10}$ \cite{Bobeth:2011nj}, which updates
\cite{Bobeth:2010wg, Bobeth:2011gi} by taking into account the new data from
2011 and additionally the branching ratio of $B\to K\,\ell^+\ell^-$.  The upper
bound on $|C_{10}|$ implies an upper bound on $Br(B_s \to \mu^+\mu^-) \lesssim 9
\times 10^{-9}$ \cite{Bobeth:2011gi} whereas the upper bound on $|C_9|$ can be
translated into a lower bound on the position of the zero-crossing of the
$A_{\rm FB}$ in $B\to K^*\ell^+\ell^-$ \cite{Bobeth:2011nj} in this scenario. At
high-$q^2$, $B \to K \ell^+\ell^-$ offers a second observable $F_H^\ell$
\cite{Bobeth:2007dw} which becomes large for $\ell = \tau$ and is sensitive to
$C_{10}$ \cite{Bobeth:2011nj}.

In the absence of strong direct $b\to s\, \tau^+\tau^-$ constraints, a global
fit of $b\to s\,\ell^+\ell^-$ data combined with $b\to s\,\gamma$ proved also
useful to provide indirect constraints due to operator mixing on $b\to s\,
\tau^+\tau^-$ operators. A model-independent study of absorptive BSM
contributions to $\Gamma_{12}^s$ in $B_s$-mixing due to $b\to s\,\tau^+\tau^-$
showed that they do not exceed 40\% deviation from the SM prediction
\cite{Bobeth:2011st}.

This summer CDF has presented the first measurement of the transversity
observable $A_T^{(2)}$ and the observable $A_{im} \sim J_9/(d\Gamma/dq^2)$
\cite{Aaltonen:2011qs} which appear in the single-differential angular
distribution w.r.t. the angle $\phi$ in $B\to K^{*}(\to K\pi)\, \ell^+ \ell^-$.
This completes the other two previously measured single-differential
distribution w.r.t. to $\cos\theta_\ell$ and $\cos\theta_{K^*}$ which allowed to
determine $A_{\rm FB}$ and $F_L$. As already emphasized, especially LHCb will
collect larger data sets, hopefully enabling the full angular analysis
w.r.t. all 3 angles and a smaller $q^2$-binning. The measurement of the angular
observables $J_i$ and the specially designed observables $A_T^{(2,3,4,5,{\rm re,
    im})}$ and $H_T^{(2,3,4,5)}$ with reduced hadronic uncertainties would be a
welcome input of a global analysis of $b\to s\,\ell^+\ell^-$. On the theoretical
side, especially form factor predictions at high-$q^2$ are not fully available
yet \cite{Liu:2011ra} and moreover a better understanding of the sub-leading
corrections in the power expansion can help to improve the confidence in the
uncertainty estimates.

\vskip-0.0cm

\noindent {\bf Acknowledgments} $\quad$ I am indebted to the organisers of the
{\em EPS-HEP 2011} for the opportunity to present a talk and the kind
hospitality in Grenoble.  I thank Frederik Beaujean, Uli Haisch, Gudrun Hiller,
Danny van Dyk and Christian Wacker for our fruitful collaboration.



\end{document}